\documentclass[prd,superscriptaddress]{revtex4}
\usepackage{CJK}
\usepackage{amsmath}
\usepackage{enumerate}
\usepackage{amssymb}
\usepackage{amsfonts}
\usepackage{mathrsfs}
\usepackage{latexsym}
\usepackage{bm}
\usepackage{amscd}
\usepackage{graphicx}
\usepackage{epsfig}
\usepackage{hhline,multirow}
\usepackage{dcolumn}
\usepackage{url}
\usepackage[colorlinks=true,linkcolor=blue]{hyperref}
\usepackage{color}
\usepackage{indentfirst}
\usepackage{subfigure}
\usepackage{psfrag}
\usepackage{slashed}
\usepackage{float}
\usepackage{setspace}

\begin{document}

\title{Nucleon electromagnetic form factors with non-local chiral effective Lagrangian}
\author{Fangcheng He}

\affiliation{Institute of High Energy Physics, CAS, P. O. Box
918(4), Beijing 100049, China}
\affiliation{University of Chinese Academy of Sciences, Beijing 100049, China}
\author{P. Wang}

\affiliation{Institute of High Energy Physics, CAS, P. O. Box
918(4), Beijing 100049, China}
\affiliation{Theoretical Physics Center for Science Facilities,
CAS, Beijing 100049, China}

\begin{abstract}

The relativistic version of finite-range-regularisation is proposed. The covariant regulator is generated from
the nonlocal Lagrangian. This nonlocal interaction is gauge invariant and is applied to study the nucleon electromagnetic 
form factors at momentum transfer up to 2 GeV$^2$. Both octet and decuplet intermediate states are included
in the one loop calculation. Using a dipole regulator with $\Lambda$ around 0.85 GeV, the obtained form factors, 
electromagnetic radii as well as the ratios of the form factors are all comparable with the experimental data. This successful
application of chiral effective Lagrangian to relatively large momentum transfer make it possible to further investigation
of hadron quantities at high $Q^2$.

\end{abstract}


\maketitle

\section{Introduction}
The study of the properties of hadrons continues to attract significant interest in the process of 
revealing and understanding the essential mechanisms of the strong interactions. The investigation 
of the electromagnetic form factors of nucleon is very important to help us discover their internal structure. 
Though QCD is the fundamental theory to describe strong interactions, it is difficult to study 
hadron physics using QCD directly. There are many phenomenological models, such as the cloudy bag 
model \cite{Lu:1997sd}, the constituent quark model \cite{Berger:2004yi,JuliaDiaz:2003gq}, 
the $1/N_c$ expansion approach \cite{Buchmann:2002et}, the perturbative chiral quark model \cite{Cheedket:2002ik}, 
the extended vector meson dominance model \cite{Williams:1996id}, the SU(3) chiral quark model \cite{Shen:1997jd}, 
the quark-diquark model \cite{Jakob:1993th,Hellstern:1995ri}, etc. 

Besides the phenomenological models, there are also many lattice-QCD calculations for the electromagnetic 
form factors \cite{Zanotti:2003gc, Boinepalli:2006xd, Alexandrou:2006ru, Gockeler:2003ay, Gockeler:2007hj, 
Edwards:2005kw, Alexandrou:2005fh}. Lattice simulation is the most rigorous approach which starts from 
the first principles. Due to the computing limit, most quantities are still calculated with large quark ($\pi$) mass.

In hadron physics, another important method is chiral perturbation theory (ChPT). Heavy baryon and 
relativistic chiral perturbation theory have been widely applied to study the hadron spectrum and structure. 
Historically, most formulations of ChPT are based on dimensional or infrared regularisation. Though ChPT 
is a successful and systematic approach, for the nucleon electromagnetic form factors, it is only valid for 
$Q^2 < 0.1$ GeV$^2$ \cite{Fuchs:2003ir}.  When vector mesons are included, the result is close to the 
experiments  with $Q^2$ less than 0.4 GeV$^2$ \cite{Kubis:2000zd}. Therefore, with traditional ChPT, 
it is hard to study the form factors at relatively large $Q^2$, for example, to explain the $G_E/G_M$ puzzle at large $Q^2$.

An alternative regularization method, namely finite-range-regularization (FRR) has been proposed. 
Inspired by quark models that account for the finite-size of the nucleon as the source of the pion cloud, 
effective field theory with FRR has been widely applied to extrapolate the vector meson mass, magnetic moments, magnetic form factors, 
strange form factors, charge radii, first moments of GPDs, nucleon spin, etc \cite{Young:2002ib,Leinweber:2003dg,Wang:2007iw,Wang:2010hp,Allton:2005fb,Armour:2008ke,
Hall:2013oga,Leinweber:2004tc,Wang:1900ta,Wang:2012hj,Wang:2013cfp,Hall:2013dva,Wang:2015sdp,Li:2015exr,Li:2016ico,Wang:2008vb}. 
In the finite-range-regularization, there is no cut 
for the energy integral. The regulator is not covariant and is in 3-dimensional momentum space. 
This non-relativistic regulator can only be applied with the heavy baryon ChPT. 
A lot of investigations have been done for the finite range regularization and we have good knowledge on the non-relativistic 
regulator which was kept same for all the above calculations. But we know little about the relativistic 
regulator and we try to determine the relativistic regulator from the well-known form factors of nucleon. 

In this paper, we will provide a relativistic version of FRR. If we simply replace the non-relativistic 
regulator with a covariant one, the local gauge symmetry and charge conservation will be destroyed. 
As a result, the renormalized proton (neutron) charge is not 1 (0). Therefore, we generate the covariant 
regulator from the local gauge invariant Lagrangian. As a result, the nonlocal Lagrangian will be introduced. 
Using this nonlocal chiral effective Lagrangian, we will study the electromagnetic form factors up to 
$Q^2=2$ GeV$^2$. The paper is organized in the following way. In section II, we briefly introduce the 
chiral Lagrangian and construct the nonlocal interactions. The matrix elements of the nucleon electromagnetic 
current is derived in section III. Numerical results are presented in section IV. Finally, section V is a summary. 

\section{Chiral  Effective Lagrangian}
The lowest order chrial Lagrangian for baryons, pseudoscalar mesons and their interaction
can be written as \cite{Jenkins:1991ts,Jenkins:1992pi}.
\begin{eqnarray}
\mathcal{L} &=& i\, Tr\,\bar{B} \gamma_{\mu}\,\slashed{\mathscr{D}}B -m_B\,Tr\,\bar{B}B
+\bar{T}_\mu^{abc}(i\gamma^{\mu\nu\alpha}D_\alpha\,-\,m_T\gamma^{\mu\nu})T_\nu^{abc}
+\frac{f^2}{4}Tr\,\partial_\mu\Sigma\partial^\mu\Sigma^+ +D\,Tr\, \bar{B} \gamma_\mu \gamma_5\, \{A_\mu,B\} \nonumber \\ 
 &+& F\,Tr\, \bar{B} \gamma_\mu \gamma_5\, [A_\mu,B]
 +\left [\frac{{\cal C}}{f}\epsilon^{abc}\bar{T}_\mu^{ade}(g^{\mu\nu}+z\gamma_\mu\gamma_\nu) B_c^e\partial_\nu\phi_b^d+H.C \right ],
\end{eqnarray}
where $D$, $F$ and $\cal C$ are the coupling constants.
The chiral covariant derivative $\mathscr{D}_\mu$ is defined as $\mathscr{D}_\mu
B=\partial_\mu B+[V_\mu,B]$. The pseudoscalar meson octet
couples to the baryon field through the vector and axial vector
combinations as
\begin{equation}
V_\mu=\frac12(\zeta\partial_\mu\zeta^\dag+\zeta^\dag\partial_\mu\zeta)+\frac{1}{2}ie\mathscr{A}^\mu(\zeta^+Q\zeta+\zeta Q\zeta^+),~~~~
A_\mu=\frac12(\zeta\partial_\mu\zeta^\dag-\zeta^\dag\partial_\mu\zeta)-\frac{1}{2}e\mathscr{A}^\mu(\zeta Q\zeta^+-\zeta^+Q\zeta),
\end{equation}
where
\begin{equation}
\zeta=e^{i\phi/f}, ~~~~~~ f=93~{\rm MeV}.
\end{equation}
The matrix of pseudoscalar fields $\phi$ is expressed as
\begin{eqnarray}
\phi=\frac1{\sqrt{2}}\left(
\begin{array}{lcr}
\frac1{\sqrt{2}}\pi^0+\frac1{\sqrt{6}}\eta & \pi^+ & K^+ \\
~~\pi^- & -\frac1{\sqrt{2}}\pi^0+\frac1{\sqrt{6}}\eta & K^0 \\
~~K^- & \bar{K}^0 & -\frac2{\sqrt{6}}\eta
\end{array}
\right).
\end{eqnarray}
$\mathscr{A}^\mu$ is the photon field. 
The covariant derivative $D_\mu$ in the decuplet part is defined as
$D_\nu T_\mu^{abc} = \partial_\nu T_\mu^{abc}+(\Gamma_\nu,T_\mu)^{abc}$,  where $\Gamma_\nu$ 
is the chrial connection\cite{Scherer:2002tk} defined as $(X,T_\mu)=(X)_d^aT_\mu^{dbc}+(X)_d^bT_\mu^{adc}+(X)_d^cT_{\mu}^{abd}$. 
$\gamma^{\mu\nu\alpha}$,$\gamma^{\mu\nu}$ are the antisymmetric matrices expressed as
\begin{equation}
\gamma^{\mu\nu}
=\frac12\left[\gamma^\mu,\gamma^\nu\right]\hspace{.5cm}\text{and}\hspace{.5cm}
\gamma^{\mu\nu\rho}=\frac14\left\{\left[\gamma^\mu,\gamma^\nu\right],
\gamma^\rho\right\}\,
\end{equation}
In the chiral $SU(3)$ limit, the octet and decuplet baryons will have the same
mass $m_B$ and $m_T$. In our calculation, we use the physical masses for
baryon octets and decuplets. The explicit form of the baryon octet
is written as
\begin{eqnarray}
B=\left(
\begin{array}{lcr}
\frac1{\sqrt{2}}\Sigma^0 +\frac1{\sqrt{6}}\Lambda &
\Sigma^+ & p \\
~~\Sigma^- & -\frac1{\sqrt{2}}\Sigma^0+\frac1{\sqrt{6}}\Lambda & n \\
~~\Xi^- & \Xi^0 & -\frac2{\sqrt{6}}\Lambda
\end{array}
\right).
\end{eqnarray}
For the baryon decuplets, there are three indices, defined as
\begin{eqnarray}
T_{111}=\Delta^{++}, ~~ T_{112}=\frac1{\sqrt{3}}\Delta^+, ~~
T_{122}=\frac1{\sqrt{3}}\Delta^0, \\ \nonumber T_{222}=\Delta^-, ~~
T_{113}=\frac1{\sqrt{3}}\Sigma^{\ast,+}, ~~
T_{123}=\frac1{\sqrt{6}}\Sigma^{\ast,0}, \\ \nonumber
T_{223}=\frac1{\sqrt{3}}\Sigma^{\ast,-}, ~~
T_{133}=\frac1{\sqrt{3}}\Xi^{\ast,0}, ~~
T_{233}=\frac1{\sqrt{3}}\Xi^{\ast,-}, ~~ T_{333}=\Omega^{-}.
\end{eqnarray}
The octet, decuplet and octet-decuplet transition magnetic moment
operators are needed in the one loop calculation of nucleon electromagnetic
form factors. The baryon octet anomalous magnetic Lagrangian is written as
\begin{equation}\label{lomag}
{\cal L}=\frac{e}{4m_N}\left(c_1{\rm Tr}\bar{B} \sigma^{\mu\nu}
\left\{F^+_{\mu\nu},B\right\}+c_2{\rm Tr}\bar{B}
\sigma^{\mu\nu} \left[F^+_{\mu\nu},B \right]\right),
\end{equation}
where
\begin{equation}
F^+_{\mu\nu}=-\frac12\left(\zeta^\dag F_{\mu\nu}Q\zeta+\zeta
F_{\mu\nu}Q\zeta^\dag\right).
\end{equation}
The transition magnetic operator is
\begin{equation}
{\cal L}=i\frac{e}{4m_N}\mu_TF_{\mu\nu}(\epsilon_{ijk}Q^i_j\bar{B}^j_m\gamma^\mu\gamma_5T^{\nu,klm}+\epsilon^{ijk}Q^l_i\bar{T}^\mu_{klm}\gamma^\nu\gamma_5B^m_j),
\end{equation}
where the matrix $Q$ is defined as $Q=$diag$\{2/3,-1/3,-1/3\}$. At the lowest
order, the Lagrangian will generate the following nucleon anomalous magnetic
moments:
\begin{equation}\label{treemag}
F_2^p=\frac13 c_1+c_2,~~~~~~ F_2^n=-\frac23 c_1.
\end{equation}
In quark model, the nucleon magnetic moments can be written in terms of quark magnetic moments. For example,
$\mu_p=\frac43 \mu_u-\frac13\mu_d$, $\mu_n=\frac43 \mu_d-\frac13\mu_u$. Using $\mu_u=-2\mu_d=2\mu_s$, we can
get the following relationships
\begin{eqnarray}\label{eq:c1}
c_1=\frac32(c_2+1), ~~~~~~~~~~~c_1=\frac32\mu_u,~~~~~~~~~~~\mu_T=4c_1 
\end{eqnarray}
The effective decuplet anomalous magnetic moment operator can be  expressed as effective Lagrangian
\begin{eqnarray}\label{eq:ci}
{\cal L}=-\frac{ieF_2^T}{2m_T}\bar{T}_\mu^{abc}\sigma^{\rho\sigma}q_{\sigma}\mathscr{A}_\rho T_\mu^{abc}.
\end{eqnarray}
For each decuplet baryon, its moment $F_2^T$ can be written in terms of $c_1$. 
For example, for $\Delta^{++}$, the magnetic moment $\mu_{\Delta^{++}}\,=\,3\mu_u\,=\,2c_1$. Therefore, $F_2^{\Delta^{++}}\,=2c_1-2$.      
In our numerical calculations, the above anomalous magnetic moments of baryons at tree level which only depend on the 
parameter $c_1$ are used.

Now we construct the nonlocal Lagrangian which will generate the covariant regulator. 
The gauge invariant non-local Lagrangian can be obtained using the method in \cite{Terning:1991yt,Faessler,Wang:2014tna}. 
For instance, the local interaction including $\pi$ meson can be written as 
\begin{equation}
{\cal L_{\pi}}^{local}=\int\!\,dx \frac{D+F}{\sqrt{2}f} \bar{p}(x)\gamma^\mu\gamma_5\,n(x)(\partial_\mu+ie\,\mathscr{A}_\mu(x)) \pi^+(x).
\end{equation}
The nonlocal Lagrangian for this interaction is expressed as 
\begin{align}\label{eq:nonlocal}
{\cal L_{\pi}}^{nl}&=\int\!\,dx\int\!\,dy\frac{D+F}{\sqrt{2}f}\bar{p}(x)\gamma^\mu\gamma_5 n(x)F(x-y)
{\text {exp}[ie\int_x^y dz_\nu\int\!\,da\,\mathscr{A}^\nu(z-a)F(a)]} \nonumber \\
& \times (\partial_\mu\,+ie\int\!\,da\,\mathscr{A}_\mu(y-a)F(a)\big)\pi^+(y),
\end{align}
where $F(x)$ is the correlation function.
To guarantee the gauge invarice, the gauge link is introduced in the above Lagrangian.
The regulator can be generated automatically with correlation function.
With the same idea, the nonlocal electromagnetic interaction can also be obtained. 
For example, the local interaction between proton and photon is written as
\begin{align}
{\cal L}_{EM}^{local} = & -e \bar{p}(x) \gamma^\mu p(x) \mathscr{A}_\mu(x)
+\frac{(c_1-1)e}{4m_N} \bar{p}(x)\sigma^{\mu\nu}p(x)F_{\mu\nu}(x).
\end{align}
The corresponding nonlocal Lagrangian is expressed as
\begin{align}
 {\cal L}_{EM}^{nl} = -e \int da \bar{p}(x) \gamma^\mu p(x) \mathscr{A}_\mu(x-a)F_1(a)
+ \frac{(c_1-1)e}{4m_N}\int da \bar{p}(x)\sigma^{\mu\nu}p(x)F_{\mu\nu}(x-a)F_2(a),
\end{align} 
where $F_1(a)$ and $F_2(a)$ is the correlation function for the nonlocal electric and magnetic interactions.
The form factors at tree level which are momentum dependent can be easily obtained with the Fourier transformation.
The simplest choice is to assume that the correlation function of the nucleon electromagnetic vertex is the same 
as that of the nucleon-pion vertex, i.e. $F_1(a)=F_2(a)=F(a)$. 
Therefore, the Dirac and Pauli form factors will have the same dependence on the momentum transfer at tree level.
As a result, the obtained charge form factor of proton decreases very quickly with increasing $Q^2$ and it will become negative
after some $Q^2$. The better choice is to assume that the charge and magnetic form factors
at tree level have the same the momentum dependence as nucleon-pion vertex, i.e. $G_M^{\rm tree}(p)=c_1G_E^{\rm tree}(p) = c_1\tilde{F}(p)$,
where $\tilde{F}(p)$ is the Fourier transformation of the correlation function $F(a)$.
The corresponding function of $\tilde{F}_1(q)$ and $\tilde{F}_2(q)$ is then expressed as
\begin{eqnarray}
\tilde{F}_1^p(q)&=&\tilde{F}(q)\frac{4m_N^2+c_1Q^2}{4m_N^2+Q^2},~~~\tilde{F}_2^p(q)=\tilde{F}(q)\frac{4m_N^2}{4m_N^2+Q^2}, 
\end{eqnarray}
From the above equations, one can see that in the heavy baryon limit, these two choices are equivalent.
The nonlocal Lagrangian is invariant under the following gauge transformation 
\begin{eqnarray}
\pi^+(y)\rightarrow e^{i\alpha(y)}\pi^+(y),~~~~p(x)\rightarrow e^{i\alpha(x)}p(x), ~~~~\mathscr{A}_\mu(x)\rightarrow\,\mathscr{A}_\mu(x)-\frac1e\partial_\mu\alpha'(x),
\end{eqnarray}
where $\alpha(x)=\int\!\,da\alpha^{\prime}(x-a)F(a)$. From Eq.~(\ref{eq:nonlocal}), two kinds of couplings
between hadrons and one photon can be obtained. One is the normal interaction expressed as
\begin{equation}
{\cal L}^{nor}=ie\int\!\,dx\int\!\,dy\frac{D+F}{\sqrt{2}f}\bar{p}(x)\gamma^\mu\gamma_5 n(x)F(x-y)\pi^+(y)\int\!\,da\,\mathscr{A}_\mu(y-a)F(a),
\end{equation}
This interaction is similar as the traditional local Lagrangian except the correlation function. 
The other one is the additional interaction obtained by the expansion of the gauge link, expressed as
\begin{equation}
{\cal L}^{add}=ie\int\!\,dx\int\!\,dy\frac{D+F}{\sqrt{2}f}\bar{p}(x)\gamma^\mu\gamma_5 n(x)F(x-y)\int_x^y dz_\nu\int\!\,da\,\mathscr{A}^\nu(z-a)F(a)\partial_\mu \pi^+(y)                     
\end{equation}
The additional interaction is important to get the renormalized proton (neutron) charge 1 (0).
The Feynman rules for the nonlocal Lagrangian are listed in the Appendix.

\section{Electromagnetic Form Factors}

The Dirac and Pauli form factors are defined as
\begin{equation}\label{eq:f1f2}
<N(p')|J_\mu|N(p)>=\bar{u}(p')\left\{\gamma^\mu
F_1^N(Q^2)+\frac{i\sigma^{\mu\nu}
q_\nu}{2m_N}F_2^N(Q^2)\right\}u(p),
\end{equation}
where $q=p^\prime-p$ and $Q^2=-q^2$. $F_1^N(Q^2)$ and $F_2^N(Q^2)$ are the Dirac and Pauli form factors.
The combination of the above form factors can generate the electric and magnetic form factors as 
\begin{eqnarray}
G_E^N(Q^2)=F_1^N(Q^2)-{Q^2\over{4m_N^2}}F_2^N(Q^2)\,\,\,\,\,\,\,\,\,\,\,\,\,\, 
G_M^N(Q^2)=F_1^N(Q^2)+F_2^N(Q^2)
\end{eqnarray} 
Charge and magnetic radii are defined by
\begin{eqnarray}
<(r_E^p)^2>={-6\over{G_E^p(0)}}{dG_E^p(Q^2)\over{dQ^2}}|_{Q^2=0},\,\,\,\,\,\,\,\,\,\,\,\,\,\, 
<(r_M^p)^2>={-6\over{G_M^p(0)}}{dG_M^p(Q^2)\over{dQ^2}}|_{Q^2=0},\\ 
<(r_E^n)^2>=-6{dG_E^n(Q^2)\over{dQ^2}}|_{Q^2=0},\,\,\,\,\,\,\,\,\,\,\,\,\,\, 
<(r_M^n)^2>={-6\over{G_M^n(0)}}{dG_M^n(Q^2)\over{dQ^2}}|_{Q^2=0}.
\end{eqnarray}
According to the Lagrangian, the one loop Feynman diagrams which contribute to the nucleon
electromagnetic form factors are plotted in Fig.~1. 

\begin{figure}[tbp]
\begin{center}
\includegraphics[scale=0.85]{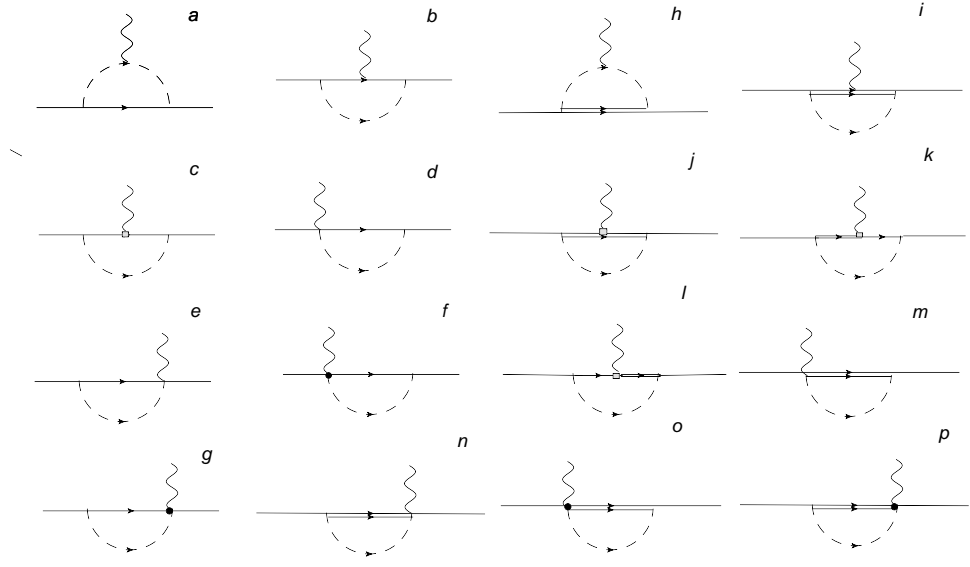}
\caption{One-loop Feynman diagrams for the nucleon electromagnetic form factors. The solid, double-solid, dashed and wave lines 
are for the octet baryons, decuplet baryons, pseudoscalar mesons and photons, respectively. The rectangle and blackdot respresent 
magentic and additional interacting vertex.}
\label{diagrams}
\end{center}
\end{figure}

In this section, we will only show the expressions for the intermediate octet baryon part. For the intermediate decuplet baryon part, 
the expressions are written in the Appendix. In diagram Fig.~1a, the photon couples to the meson. The contribution of Fig.~1a to 
the matrix element in Eq.~(\ref{eq:f1f2}) is expressed as
\begin{eqnarray} 
\Gamma_a^{\mu(p)}&=&-(D+F)^2I_{a\pi}^{N}-\frac{(3F+D)^2}{6}I_{aK}^{\Lambda}-\frac{(D-F)^2}{2}I_{aK}^{\Sigma},\\
\Gamma_a^{\mu(n)}&=&(D+F)^2I_{a\pi}^{N}-(D-F)^2I_{aK}^{\Sigma},
\end{eqnarray} 
where $I_{a\pi}^{N}$, $I_{aK}^{\Lambda}$ and $I_{aK}^{\Sigma}$ are the integrals for the
$N\pi$, $\Lambda K$ and $\Sigma K$ intermediate states, respectively. 
$I_{a\pi}^{N}$ is expressed as
\begin{equation} 
I_{a\pi}^{N}=\bar{u}(p')\int\!\frac{d^4 k}{(2\pi)^4}\frac{(\slashed{k}+\slashed{q})\gamma_5}{\sqrt{2}f}\,\tilde{F}(q+k)
\frac{1}{D_\pi(k+q)}\,(2k+q)^\mu\frac{1}{D_\pi(k)}\frac{1}{\slashed{p}-\slashed{k}-m_N}\frac{-\slashed{k}\gamma_5}{\sqrt{2}f}\tilde{F}(k)u(p).
\end{equation}
$D_\pi(k)$ is given by 
\begin{equation} 
D_\pi(k)=k^2-M_k^2+i\epsilon.
\end{equation}
The expressions for $I_{aK}^{\Lambda}$ and $I_{aK}^{\Sigma}$ are the same except the intermediate meson and baryon masses
are changed to be those of $K$ meson and hyperons. For simplisity, we will only show the expression for the $\pi$ meson case.

In Fig.1b, the photon couples to the intermediate baryon with electric vertex. 
The contribution of this diagram with octet intermediate baryons
is expressed as 
\begin{eqnarray} 
\Gamma_b^{\mu(p)}&=&{{1}\over{2}}(D+F)^2\frac{12m_p^2-c_1Q^2}{12m_p^2+3Q^2}I_{b\pi}^{NN}+{{(3F-D)^2}\over{6}}
\frac{4m_p^2+c_1Q^2}{4m_p^2+Q^2}I_{b\eta}^{NN}+(D-F)^2\frac{24m_\Sigma^2+7c_1Q^2}{24m_\Sigma^2+6Q^2}I_{bK}^{N\Sigma} \nonumber \\
&-& \frac{(3F+D)^2}{18}\frac{c_1Q^2}{4m_\Lambda^2+Q^2}I_{bK}^{N\Lambda}
-\frac{(3F+D)(D-F)}{3}\frac{c_1Q^2}{4m_\Sigma^2+Q^2}I_{bK}^{N\Lambda\Sigma},
\end{eqnarray} 
\begin{eqnarray} 
\Gamma_b^{\mu(n)}&=&(D+F)^2\frac{12m_p^2+2c_1Q^2}{12m_p^2+3Q^2}I_{b\pi}^{NN}
-{{(3F-D)^2}\over{9}}
\frac{c_1Q^2}{4m_p^2+Q^2}I_{b\eta}^{NN}
-{{(3F+D)^2}\over{18}}\frac{c_1Q^2}{4m_\Lambda^2+Q^2}I_{bK}^{N\Lambda} \nonumber\\
&-&(D-F)^2\frac{c_1Q^2+24m_\Sigma^2}{24m_\Sigma^2+6Q^2}I_{bK}^{N\Sigma}
+\frac{(3F+D)(D-F)}{3}\frac{c_1Q^2}{4m_\Sigma^2+Q^2}I_{bK}^{N\Lambda\Sigma},
\end{eqnarray} 
where the integral $I_{b\pi}^{NN}$ is written as
\begin{equation} 
I_{b\pi}^{NN}=\tilde{F}(q)\bar{u}(p')\int\!\frac{d^4 k}{(2\pi)^4}\frac{\slashed{k}\gamma_5}{\sqrt{2}f}\,\tilde{F}(k)\frac{1}{D_\pi(k)}
\frac{1}{\slashed{p'}-\slashed{k}-m_N}(-\gamma_\mu)\frac{1}{\slashed{p}-\slashed{k}-m_N}\frac{-\slashed{k}\gamma_5}{\sqrt{2}f}\tilde{F}(k)u(p).
\end{equation}

Fig.1c is for the magnetic baryon-photon interaction. The contribituon of this diagram is expressed as
\begin{eqnarray} 
\Gamma_c^{\mu(p)}&=&-\frac{(4c_1+12)m_p^2}{12m_p^2+3Q^2}(D+F)^2I_{c\pi}^{NN}+\frac{(4c_1-4)m_p^2}{12m_p^2+3Q^2}(3F-D)^2I_{c\eta}^{NN}
-\frac{4c_1m_\Lambda^2}{36m_\Lambda^2+9Q^2}(3F+D)^2 I_{cK}^{N\Lambda}        \nonumber\\
&+&\frac{(28c_1-24)m_\Sigma^2}{12m_\Sigma^2+3Q^2}(D-F)^2 I_{cK}^{N\Sigma}-\frac{8c_1(D-F)(3F+D)m_\Sigma^2}{12m_\Sigma^2+3Q^2}I_{cK}^{N\Lambda\Sigma},                            \\
\Gamma_c^{\mu(n)}&=&\frac{(16c_1-24)m_p^2}{12m_p^2+3Q^2}(D+F)^2 I_{c\pi}^{NN}-\frac{8c_1m_N^2}{36m_N^2+9Q^2}(3F-D)^2 I_{c\eta}^{NN}
-\frac{4c_1 m_\Lambda^2}{36m_\Lambda^2+9Q^2}(3F+D)^2 I_{cK}^{N\Lambda}                            \nonumber\\
&-&\frac{(4c_1-24)m_\Sigma^2}{12m_\Sigma^2+3Q^2} (D-F)^2 I_{cK}^{N\Sigma}
+\frac{8c_1(D-F)(3F+D)m_\Sigma^2}{12m_\Sigma^2+3Q^2}I_{cK}^{N\Lambda\Sigma},
\end{eqnarray} 
where
\begin{eqnarray}
I_{c\pi}^{NN}&=&\tilde{F}(q)\bar{u}(p')\int\!\frac{d^4 k}{(2\pi)^4}\frac{\slashed{k}\gamma_5}{2f}\tilde{F}(k)\frac{1}{\slashed{p'}-\slashed{k}-m_N}
\frac{\sigma^{\mu\nu}q_{\nu}}{2m_N}\frac{1}{\slashed{p}-\slashed{k}-m_N}\frac{i}{D_\pi(k)}\frac{\slashed{k}\gamma_5}{2f}\tilde{F}(k)u(p).\\
\end{eqnarray}

The contribution from Fig.~1d+1e is written as 
\begin{eqnarray} 
\Gamma_{d+e}^{\mu(p)}&=&-(D+F)^2I_{(d+e)\pi}^{NN}-\frac{(3F+D)^2}{6}I_{(d+e)K}^{N\Lambda}-\frac{(D-F)^2}{2}I_{(d+e)K}^{N\Sigma},\\
\Gamma_{d+e}^{\mu(n)}&=&(D+F)^2I_{(d+e)\pi}^{NN}-(D-F)^2I_{(d+e)K}^{N\Sigma},
\end{eqnarray} 
where
\begin{eqnarray}
I_{(d+e)\pi}^{NN}&=&\tilde{F}(q)\bar{u}(p')\int\!\frac{d^4 k}{(2\pi)^4}\frac{\slashed{k}\gamma_5}{\sqrt{2}f}\tilde{F}(k)
\frac{1}{\slashed{p'}-\slashed{k}-m}\frac{1}{D_\pi(k)}\frac{-1}{\sqrt{2}f}\gamma^\mu\gamma_5 \tilde{F}(q-k)u(p)     \nonumber\\
&+&\tilde{F}(q)\,\bar{u}(p')\int\!\frac{d^4 k}{(2\pi)^4}\frac{1}{\sqrt{2}f}\gamma^\mu\gamma_5\tilde{F}(q+k)\frac{1}{\slashed{p}-\slashed{k}-m}
\frac{1}{D_\pi(k)}\frac{-\slashed{k}\gamma_5}{\sqrt{2}f}\tilde{F}(k)u(p).
\end{eqnarray} 
These two diagrams only have contribution in the relativistic cases. In the heavy baryon limit, 
they have no contribution to either electric or magnetic form factors.

Fig.~1f and 1g are the additional diagrams which generated from the expansion of the gauge link terms.
They are important to get the renormalized charge to proton (neutron) to be 1 (0).
The contribution of these two additional diagrams with intermediate octet baryons is expressed as
\begin{eqnarray} 
\Gamma_{f+g}^{\mu(p)}&=&-(D+F)^2I_{(f+g)\pi}^{NN}-\frac{(3F+D)^2}{6}I_{(f+g)K}^{N\Lambda}-\frac{(D-F)^2}{2}I_{(f+g)K}^{N\Sigma},\\
\Gamma_{f+g}^{\mu(n)}&=&(D+F)^2I_{(f+g)\pi}^{NN}-(D-F)^2I_{(f+g)K}^{N\Sigma},
\end{eqnarray} 
where
\begin{eqnarray}
I_{(f+g)\pi}^{NN}&=&\tilde{F}(q)\,\bar{u}(p')\int\!\frac{d^4 k}{(2\pi)^4}\frac{\slashed{k}\gamma_5}{\sqrt{2}f}\tilde{F}(k)\frac{1}{\slashed{p'}
-\slashed{k}-m}\frac{1}{D_\pi(k)}\frac{1}{\sqrt{2}f}(-\slashed{k}\gamma_5)\frac{(-2k+q)^\mu}{-2kq+q^2}[\tilde{F}(k-q)-\tilde{F}(k)]u(p) \nonumber\\
&+&\tilde{F}(q)\,\bar{u}(p')\int\!\frac{d^4 k}{(2\pi)^4}\frac{1}{\sqrt{2}f}\slashed{k}\gamma_5\frac{(2k+q)^\mu}{2kq+q^2}[\tilde{F}(k+q)-\tilde{F}(k)]
\frac{1}{\slashed{p}-\slashed{k}-m}\frac{1}{D_\pi(k)}\frac{\slashed{k}\gamma_5}{\sqrt{2}f}\tilde{F}(k)u(p).
\end{eqnarray} 
Using FeynCalc to simplify the $\gamma$ matrix algebra, we can get the separate expressions for the Dirac and Pauli form factors.
Numerical results will be discussed in the next section.

\begin{figure}[tbp]
\begin{center}
\includegraphics[scale=0.85]{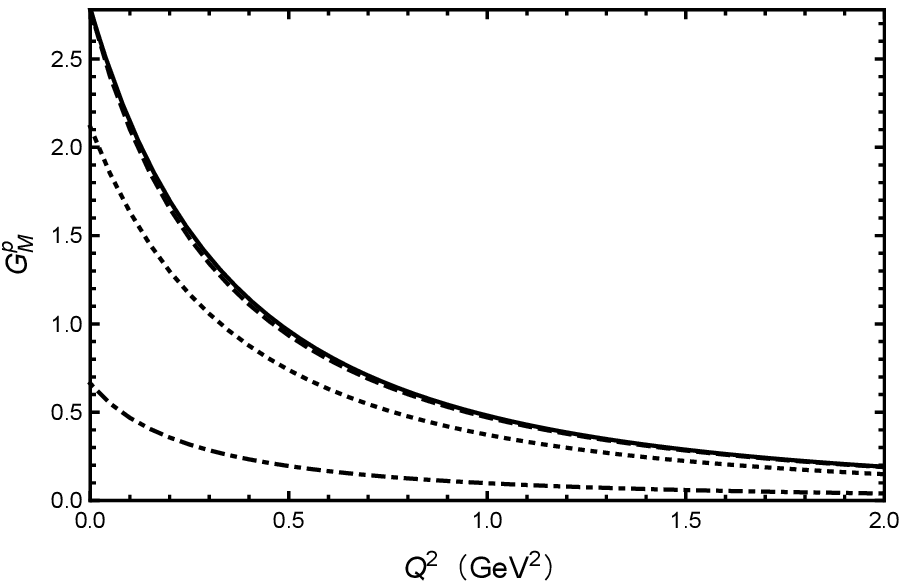}
\caption{The proton magnetic form factor versus momentum transfer $Q^2$ with $\Lambda=0.85$ GeV. 
The solid line is for the empirical result. The dotted, dot-dashed and dashed lines are for the tree, loop and 
total contribution, respectively.}
\label{diagrams}
\end{center}
\end{figure}

\begin{figure}[t]
\begin{center}
\includegraphics[scale=0.85]{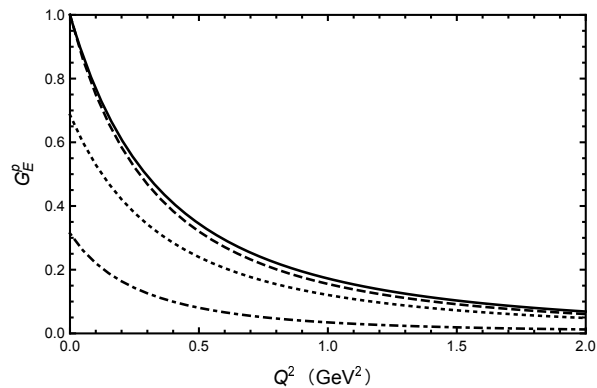}
\caption{Same as Fig.~2 but for the proton electric form factor.}
\label{diagrams}
\end{center}
\end{figure}

\section{Numerical Results}

In the numerical calculations, the parameters are chosen as $D=0.76$
and $F=0.50$ ($g_A=D+F=1.26$). The coupling constant ${\cal C}$ is
chosen to be $1$ which is the same as in Ref.~\cite{Pascalutsa:2006up}.
The off-shell parameter $z$ is chosen to be $z=-1$ \cite{Nath:1971wp}. The low energy constant $c_1$ is fitted by the
experimental moment of $F_2^n(0) = -1.91$.
The covariant regulator is chosen to be of a dipole form
\begin{equation}
\tilde{F}(k)=\frac{1}{(1-k^2/\Lambda^2)^2},
\end{equation} 
where Lambda is the only free parameter. By varying the value of $\Lambda$, we found when $\Lambda$ is arond 0.85 GeV, 
the results are very close to the experimental nucleon form factors.

\begin{figure}[t]
\begin{center}
\includegraphics[scale=0.9]{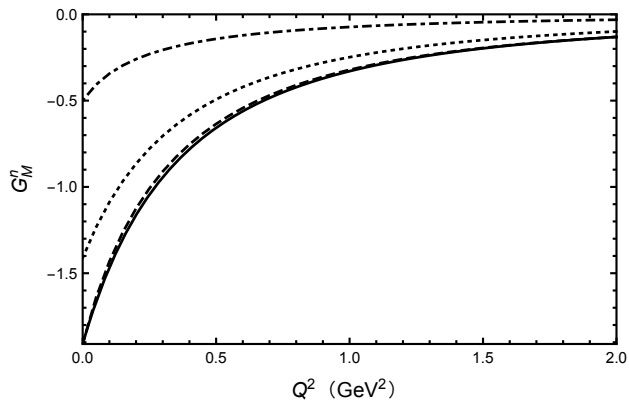}
\caption{The magnetic form factor of neutron versus momentum transfer $Q^2$ with $\Lambda=0.85$ GeV. 
The solid line is for the empirical result. The dotted, dot-dashed and dashed lines are for the tree, loop and 
total contribution, respectively.}
\label{diagrams}
\end{center}
\end{figure}

\begin{figure}[t]
\begin{center}
\includegraphics[scale=0.85]{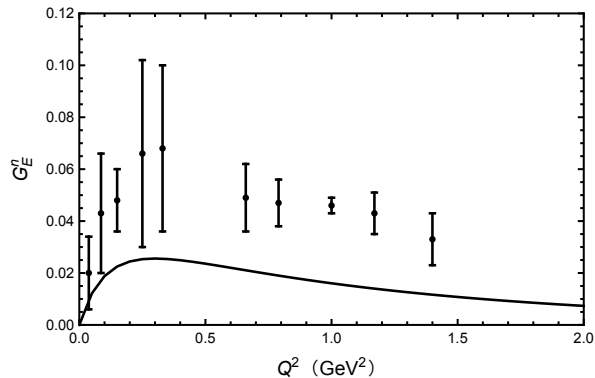}
\caption{The electric form factor of neutron versus momentum transfer $Q^2$ with $\Lambda=0.85$ GeV. 
The experimental date is from \cite{Seimetz:2005vg}.}
\label{diagrams}
\end{center}
\end{figure}

The calculated proton magnetic form factor $G_M^p(Q^2)$ versus $Q^2$ is plotted in Fig.~2. 
The solid line is for the empirical result with $G^p_M(Q^2) = 2.79/(1 +Q^2/0.71$ GeV$^2)^2$. 
The dotted, dot-dashed and dashed lines are for the tree, loop and total contribution, respectively. 
As we explained previously, on the one hand, the nonlocal Lagrangian generates the covariant regulator 
which makes the loop integral convergent. On the other hand, it also generates the  
$Q^2$ dependent contribution at tree level. Compared with the conventional ChPT, the tree level
contribution is not expanded in powers of momentum transfer. 
As a result, both the tree and loop contribution decrease smoothly with the increasing $Q^2$ 
and the total obtained form factor is
close to the experimantal value up to $Q^2 = 2$ GeV$^2$. For $Q^2=0$, the contribution to $\mu^p$
at tree level is 2.11 and the loop contribution to $\mu^p$ is 0.67. The total $\mu^p$ is 2.78.
This proton magnetic moment is calculated with fixed $c_1$ which is determined by the neutron
magnetic moment ($\mu^n = -1.91$). The proton magnetic radii is 0.848 fm in our calculation, 
which is obviously close to the experimental value.

The proton charge form factor versus $Q^2$ is shown in Fig~3. The solid, dashed,
dotted and dot-dashed lines have the same meaning as Fig.~2 except for the charge form factor.
From the figure, one can see both the tree and loop contribution are important to get
the correct $Q^2$ dependence of the form factors. At $Q^2=0$, the sum of the tree and loop
contribution to proton charge is 1. The additional diagrams generated from the expansion of 
the gauge link is crucial to get the renormalized proton charge 1. Compared with the 
magnetic form factor, the charge form factor decreases faster. As a result, the obtained
charge radii 0.857 fm is a little larger than the magnetic radii 

The neutron magnetic form factor versus $Q^2$ is shown in Fig.~4. Similar as the proton case, the solid line is for the empirical result. 
The dotted, dot-dashed and dashed lines represent the tree, loop and total contribution to the neutron form factor, respectively. 
Again, compared with the empirical data, our calculated result is very good up to $Q^2=2$ GeV$^2$. 
The calculated magnetic radii of neutron is 0.867 fm. From Fig.~2 to Fig.~4, we can see the loop diagrams contribute about $25\%-30\%$ 
to proton electromagnetic form factors and neutron magnetic form factor, while $70\%-75\%$ of the form factors is from the 
tree level contribution.

The neutron charge form factor is plotted in Fig~5. Since the charge of neutron is 0, all the contribution to
the neutron charge form factor is from the loop. It first increases and then decreases with the increasing
momentum transfer. The neutron charge radii $<(r_M^n)^2>=-0,077$ fm$^2$, which is smaller than experimental
value $-0.11$ fm$^2$. Though the calculated charge form factor of neutron is smaller than experimental values, 
overall the result is still reasonable.  

\begin{figure}[H]
\begin{center}
\includegraphics[scale=0.9]{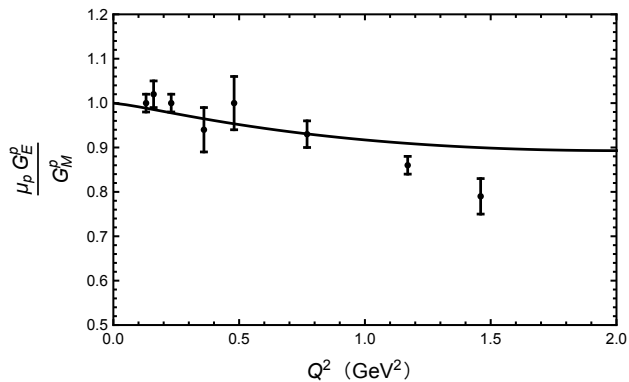}
\caption{Radio of proton electric to normalized magnetic form factor versus momentum transfer $Q^2$. 
Experimental result is from \cite{Ostrick:2006jd}}
\label{diagrams}
\end{center}
\end{figure}

\begin{figure}[H]
\begin{center}
\includegraphics[scale=0.9]{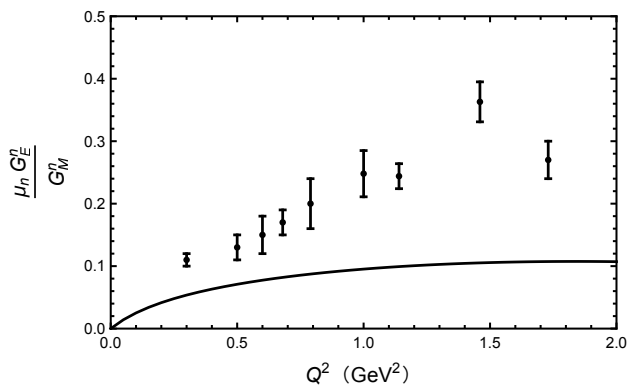}
\caption{Radio of neutron electric to normalized magnetic form factor versus momentum transfer $Q^2$. 
Experimental result is from \cite{Riordan:2010id}}
\label{diagrams}
\end{center}
\end{figure}

In the traditional ChPT, in addition to the two parameters
$c_1$ and $c_2$ which were determined by the proton and neutron magnetic moments, there are 
four other parameters fitted by the electric and magnetic radii of proton and neutron.
Here besides the parameter $c_1$ fitted by the exprimental neutron magnetic moment,
we have only one free parameter $\Lambda$ in the regulator.
The proton magnetic moment and the nucleon radii are calculated instead of fitted.
With fewer parameters, the obtained electromagnetic form factors of proton and neutron are all much better than 
those in the traditional ChPT. This makes it possible to study the form factors precisely at relatively large $Q^2$.

With the precisely determined form factors, we now show the ratios of the electric to normalized magnetic form factor. 
The ratio for proton is plotted in Fig~6. If without loop contribution, the ratio will remain to be 1 for all $Q^2$. 
With loop contribution, $\frac{\mu_pG_E^p}{G_M^p}$ automatically deceases with the increasing $Q^2$. 
Our calculated result is comparable with the experimental data, though at large $Q^2$, the experimental data drop more quickly.

The ratio for neutron is plotted in Fig~7. From the figure, one can see the radio $\frac{\mu_nG_E^n}{G_M^n}$ 
increases with the increasing $Q^2$ as the experimental data. This is purely due to the loop contribution.
The experimental ratio of $\frac{\mu_nG_E^n}{G_M^n}$ increases more quickly than our result. It is mainly because 
our calculated $G_E^n$ is smaller than the experimental data.

\section{Summary}

We proposed a relativistic version for the finite-range-regularization which makes it possible to
study the hadron properties with relativistic chiral effective Lagrangian at large $Q^2$.
The finite-range-regularization has been widely applied to investigate the nucleon mass, form factors,
electromagnetic radii, generalized parton distributions, proton spin, etc. We have good knowledge
on the 3-dimensional regulator which was kept the same for all the calculations. However, we have little
knowledge on the covariant 4-dimensional regulator. Therefore, we start from the well-determined nucleon form
factors and it was found that using the dipole regulator with $\Lambda$ around 0.85 GeV
the nucleon form factors can be described very well up to $Q^2 =2$ GeV$^2$.
The covariant regulator is generated from the nonlocal gauge invariant Lagrangian. As a result,
the renomalized charge of proton (neutron) is 1 (0) with the additional diagrams obtained by the 
expansion of the gauge link. The nonlocal interaction generates both the regulator which makes the loop
integral convergent and the $Q^2$ dependence of form factors at tree level.
In this approach, we have only two parameters $c_1$ and $\Lambda$ instead of six parameters in
the traditional ChPT. With fewer parameters, our calculated form factors are much better.
The ratios of the electric to normalized magnetic form factor are also comparable with the experimental data.
From our calculation, the $G_N^E/G_N^M$ puzzle can be naturely understood.
This is the first time to calculate the form factors precisely at relatively large $Q^2$ with chiral effective Lagrangian.
The successful application of chiral effective Lagrangian to large momentum transfer will be very helpful for us to
investigate hadron quantities at high $Q^2$.
As a summary, we list the parameters and obtained magnetic moments and electromagnetic radii in Table I.

\begin{table}
\caption{The parameters and calculated magnetic moments and electromagnetic radii of nucleon}
\begin{ruledtabular}
\begin{tabular}{ccc|cccccccc}
 $\Lambda$ (GeV)& $Z$ &$c_1$  &$\mu_p$&$\mu_n$& $r_{Mp}$ $(fm)$ &$r_{Ep}$ $(fm)$&$r_{Mn}$ $(fm)$&$r_{En}^2$ $(fm^2)$                                                                                        \\  \hline
0.8 & 0.71 & 3.090  & 2.78 &$-1.91$ &  0.893 & 0.903 & 0.912 & -0.076   \\
0.85 & 0.69 &   3.085  & 2.78 &$-1.91$ &  0.848 & 0.857 & 0.867 & -0.077  \\
0.9 & 0.66 & 3.077 & 2.78  &$-1.91$ &  0.808 & 0.816 & 0.829 & -0.082   \\
Exp. & -& - & 2.79 & $-1.91$ &0.836 &0.847 &0.889 & -0.113 
\end{tabular}
\end{ruledtabular}
\end{table}

\section*{Acknowledgments}

This work is supported by NSFC under Grant No. 11475186, by DFG and NSFC (CRC 110) and 
by Key Research Program of Frontier Sciences, CAS under Grant NO. Y7292610K1.
\section*{Appendix}

\begin{figure}[tbp]
\begin{center}
\includegraphics[scale=0.8]{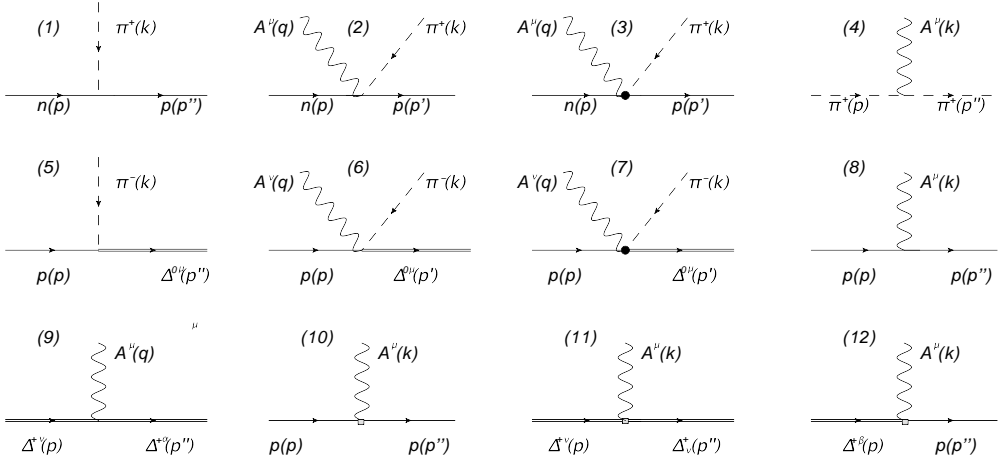}
\caption{The interacting vertex in the calculation of nucleon form factors. Only the $\pi$ case is shown as an example.
The rectangle and black dot represent the magnetic and additional interacting vertex.}
\label{diagrams1}
\end{center}
\end{figure}
The Feynman rules for the nonlocal vertexes are written as
\begin{eqnarray*}
&(1)&:\frac{\slashed{k}\gamma_5}{\sqrt{2}f}(D+F)\tilde{F}(k)    \\
&(2)&:-\frac{e}{\sqrt{2}f}(D+F)\gamma^\mu\gamma_5 \tilde{F}(k+q)\tilde{F}(q) \\
&(3)&:-\frac{e}{\sqrt{2}f}\slashed{k}\gamma_5\frac{(2k+q)^\mu}{2kq+q^2}[\tilde{F}(k+q)-\tilde{F}(k)]\tilde{F}(q) \\
&(4)&:-ie(p+p^{\prime\prime})^\mu \tilde{F}(q)\\
&(5)&:-\frac{{\cal C}}{\sqrt{6}f}\big(k_\mu+z\gamma_\mu\slashed{k}\big)\tilde{F}(k)  \\
&(6)&:-\frac{{e\cal C}}{\sqrt{6}f}(g^{\nu\mu}+z\gamma^\mu\gamma^\nu)\tilde{F}(k+q)\tilde{F}(q)                                                                                   \\
&(7)&:-\frac{{e\cal C}}{\sqrt{6}f}\big(k_\mu+z\gamma_\mu\slashed{k}\big)\frac{(2k+q)^\mu}{2kq+q^2}[\tilde{F}(k+q)-\tilde{F}(k)]\tilde{F}(q) \\
&(8)&:-ie\gamma^\mu\frac{4m_N^2+c_1Q^2}{4m_N^2+Q^2}\tilde{F}(q) \\
&(9)&:-ie\gamma^{\alpha\nu\mu}\frac{4m_\Delta^2+c_1Q^2}{4m_\Delta^2+Q^2}\tilde{F}(q)    \\
&(10)&:\frac{4e(c_1-1)m_N^2}{4m_N^2+Q^2}\frac{\sigma^{\mu\nu}q_\nu}{2m_N}\tilde{F}(q)  \\
&(11)&:-\frac{4e\,(c_1-1)m_\Delta^2}{4m_\Delta^2+Q^2}\frac{\sigma^{\mu\lambda}q_\lambda}{2m_\Delta}\tilde{F}(q)     \\
&(12)&: e\frac{\sqrt{3}}{3m_N}c_1F_{\mu\nu}\gamma^\mu\gamma_5\tilde{F}(q)
\end{eqnarray*}

The expressions for the decuplet part are written in the following way.
The contribution of Fig.~1h is expressed as
\begin{eqnarray} 
\Gamma_h^{\mu(p)}&=& \frac{2{\cal C}^2}{3}I_{h\pi}^{N\Delta}-\frac{{\cal C}^2}{6}I_{hK}^{N\Sigma^*},\\
\Gamma_h^{\mu(n)}&=& -\frac{2{\cal C}^2}{3}I_{h\pi}^{N\Delta}-\frac{{\cal C}^2}{3}I_{hK}^{N\Sigma^*},
\end{eqnarray} 
where
\begin{eqnarray} 
I_{h\pi}^{N\Delta}&=&\bar{u}(p')\int\!\frac{d^4 k}{(2\pi)^4}\frac{1}{2f^2}\big((k+q)_\sigma+z(\slashed{k}+\slashed{q})\gamma_\sigma)\big)F(q+k)\frac{1}{D_\pi(k+q)}            \nonumber\\
&\times&\,e(2k+q)^\mu\frac{1}{D_\pi(k)}\frac{1}{\slashed{p}-\slashed{k}-m_\Delta}S_{\sigma\rho}(-k_\rho-z\gamma_\rho\slashed{k})F(k)u(p).
\end{eqnarray} 
$S_{\sigma\rho}$ is expressed as
\begin{equation} 
S_{\sigma\rho}=-g_{\sigma\rho}+\frac{\gamma_\sigma\gamma_\rho}{3}+\frac{2(p-k)_\sigma(p-k)_\rho}{3m_\Delta^2}+\frac{\gamma_\sigma(p-k)_\rho-\gamma_\rho(p-k)_\sigma}{3m_\Delta}.
\end{equation}
The contribution of Fig.~1i is expressed as 
\begin{equation} 
\Gamma_i^{\mu(p)}=\frac{4 {\cal C}^2}{3}\frac{4m_\Delta^2+c_1Q^2}{4m_\Delta^2+Q^2}I_{i\pi}^{N\Delta}
+\frac{{\cal C}^2}{6}\frac{4m_\Sigma^{*2}+c_1Q^2}{4m_\Sigma^{*2}+Q^2}I_{iK}^{N\Sigma^*},
\end{equation}
\begin{equation} 
\Gamma_i^{\mu(n)}= -\frac{{\cal C}^2}{3}\frac{4m_\Delta^2+c_1Q^2}{4m_\Delta^2+Q^2}I_{i\pi}^{N\Delta}
-\frac{{\cal C}^2}{6}\frac{4m_\Sigma^{*2}+c_1Q^2}{4m_\Sigma^{*2}+Q^2}I_{iK}^{N\Sigma^*},
\end{equation} 
where
\begin{eqnarray} 
I_{i\pi}^{N\Delta}&=&\bar{u}(p')\int\!\frac{d^4 k}{(2\pi)^4}\frac{1}{2f^2}(k_\sigma+z\slashed{k}\gamma_\sigma)F(k)\frac{1}{D_\pi(k)}
\frac{1}{\slashed{p'}-\slashed{k}-m_\Delta}S_{\sigma\alpha}    \nonumber\\
&\times&(-2\gamma^{\alpha\beta\mu})\frac{1}{\slashed{p}-\slashed{k}-m_\Delta}S_{\beta\rho}(-k_\rho-z\gamma_\rho\slashed{k})F(k)u(p).
\end{eqnarray} 
The contribution of Fig.~1j is expressed as
\begin{eqnarray} 
\Gamma_j^{\mu(p)}&=& \frac{4{\cal C}^2}{3}I_{j\pi}^{N\Delta}+\frac{{\cal C}^2}{6}I_{jK}^{N\Sigma^*},\\
\Gamma_j^{\mu(n)}&=& -\frac{{\cal C}^2}{3}I_{j\pi}^{N\Delta}-\frac{{\cal C}^2}{6}I_{jK}^{N\Sigma^*},
\end{eqnarray} 
where
\begin{eqnarray} 
I_{j\pi}^{N\Delta}&=&\bar{u}(p')\int\!\frac{d^4 k}{(2\pi)^4}\frac{1}{2f^2}(k_\sigma+z\slashed{k}\gamma_\sigma)F(k)
\frac{i}{D_\pi(k)}\frac{i}{\slashed{p'}-\slashed{k}-m_\Delta}S_{\sigma\nu}\frac{(1-c_1)}{m_\Delta}\sigma^{\mu\lambda}q_\lambda \nonumber\\
 &\times&\frac{i}{\slashed{p}-\slashed{k}-m_\Delta}S_{\nu\rho}(-k_\rho-z\gamma_\rho\slashed{k})F(k)u(p).
\end{eqnarray}
The contribution of Fig.1k+1l is expressed as
\begin{eqnarray} 
\Gamma_{k+l}^{\mu(p)}&=& 2(D+F){\cal C}I_{(k+l)\pi}^{N\Delta}+ \frac{5}{4}(D-F){\cal C}I_{(k+l)K}^{\Sigma\Sigma^*}
+\frac{1}{4}(3F+D){\cal C}I_{(k+l)K}^{\Lambda\Sigma^*},\\
\Gamma_{k+l}^{\mu(n)}&=&-2(D+F){\cal C}I_{(k+l)\pi}^{N\Delta}+\frac{1}{4}(D+F){\cal C}I_{(k+l)K}^{\Sigma\Sigma^*}
-\frac{1}{4}(3F+D){\cal C}I_{(k+l)K}^{\Lambda\Sigma^*},
\end{eqnarray} 
where
\begin{align}
I_{(k+l)\pi}^{\Sigma\Sigma^*}=&-&F(q)\,\bar{u}(p')\int\!\frac{d^4 k}{(2\pi)^4}\frac{c_1}{6m_{\Sigma^*}f^2}F^2(k)\slashed{k}\gamma_5\frac{1}{\slashed{p'}-\slashed{k}-m_\Sigma}\gamma^\nu\gamma_5q_\nu\frac{1}{\slashed{p}-\slashed{k}-m_{\Sigma^*}}S_{\mu\rho}(k_\rho+z\gamma_\rho\slashed{k})\frac{i}{D_\pi(k)}u(p)              \nonumber\\
&+&F(q)\,\bar{u}(p')\int\!\frac{d^4 k}{(2\pi)^4}\frac{c_1}{6m_{\Sigma^*}f^2}F^2(k)\slashed{k}\gamma_5\frac{1}{\slashed{p'}-\slashed{k}-m_\Sigma}\gamma^\mu\gamma_5q_\nu\frac{1}{\slashed{p}-\slashed{k}-m_{\Sigma^*}}S_{\nu\rho}(k_\rho+z\gamma_\rho\slashed{k})\frac{i}{D_\pi(k)}u(p)              \nonumber\\
&-&F(q)\,\bar{u}(p')\int\!\frac{d^4 k}{(2\pi)^4}\frac{c_1}{6m_{\Sigma^*}f^2}F^2(k)(k_\nu+z\slashed{k}\gamma_\nu)\frac{1}{\slashed{p'}-\slashed{k}-m_\Sigma}S_{\nu\rho}q_\rho\gamma^\mu\gamma_5\frac{1}{\slashed{p}-\slashed{k}-m_{\Sigma^*}}\slashed{k}\gamma_5\frac{1}{D_\pi(k)}u(p)\nonumber\\
&+&F(q)\,\bar{u}(p')\int\!\frac{d^4 k}{(2\pi)^4}\frac{c_1}{6m_{\Sigma^*}f^2}F^2(k)(k_\nu+z\slashed{k}\gamma_\nu)\frac{1}{\slashed{p'}-\slashed{k}-m_\Sigma}S_{\nu\mu}q_\rho\gamma^\rho\gamma_5\frac{1}{\slashed{p}-\slashed{k}-m_{\Sigma^*}}\slashed{k}\gamma_5\frac{1}{D_\pi(k)}u(p).
\end{align}
The contribution of Fig.~1m+1n is expressed as 
\begin{eqnarray} 
\Gamma_{m+n}^{\mu(p)}&=& \frac{2{\cal C}^2}{3}I_{(m+n)\pi}^{N\Delta}-\frac{{\cal C}^2}{6}I_{(m+n)K}^{N\Sigma^*},\\
\Gamma_{m+n}^{\mu(n)}&=&-\frac{2{\cal C}^2}{3}I_{(m+n)\pi}^{N\Delta}-\frac{{\cal C}^2}{3}I_{(m+n)K}^{N\Sigma^*},
\end{eqnarray}
where
\begin{eqnarray}
I_{(m+n)\pi}^{N\Delta}&=&eF(q)\,\bar{u}(p')\int\!\frac{d^4 k}{(2\pi)^4}\frac{1}{2f^2}(k_\sigma+z\slashed{k}\gamma_\sigma)F(k)\frac{1}{D_\pi(k)}\frac{1}{\slashed{p'}-\slashed{k}-m_\Delta}S_{\sigma\rho}(g^{\rho\mu}+z\gamma^\rho\gamma^\mu)F(-k+q)u(p)                          \nonumber\\
&+&eF(q)\,\bar{u}(p')\int\!\frac{d^4 k}{(2\pi)^4}\frac{1}{2f^2}(g^{\sigma\mu}+z\gamma^\mu\gamma^\sigma)F(k+q)\frac{1}{D_\pi(k)}\frac{1}{\slashed{p}-\slashed{k}-m_\Delta}S_{\sigma\rho}F(k)(k_\rho+z\gamma_\rho\slashed{k})u(p).
\end{eqnarray}

The  contribution of Fig.~1o+1p is expressed as
\begin{eqnarray} 
\Gamma_{o+p}^{\mu(p)}&=& \frac{2{\cal C}^2}{3}I_{(o+p)\pi}^{N\Delta}-\frac{{\cal C}^2}{6}I_{(o+p)K}^{N\Sigma^*},\\
\Gamma_{o+p}^{\mu(n)}&=&-\frac{2{\cal C}^2}{3}I_{(o+p)\pi}^{N\Delta}-\frac{{\cal C}^2}{3}I_{(o+p)K}^{N\Sigma^*},
\end{eqnarray} 
where
\begin{eqnarray}
I_{(o+p)\pi}^{N\Delta}&=&-eF(q)\,\bar{u}(p')\int\!\frac{d^4 k}{(2\pi)^4}\frac{1}{2f^2}(k_\sigma+z\slashed{k}\gamma_\sigma)F(k)
\frac{1}{D_\pi(k)}\frac{1}{\slashed{p'}-\slashed{k}-m_\Delta}S_{\sigma\rho}(k_\rho+z\gamma_\rho\slashed{k})   \nonumber\\
&\times&\frac{(-2k+q)^\mu}{-2kq+q^2}[F(k-q)-F(k)]u(p)                \nonumber\\
&+&eF(q)\,\bar{u}(p')\int\!\frac{d^4 k}{(2\pi)^4}\frac{1}{2f^2}(k_\sigma+z\slashed{k}\gamma_\sigma)\frac{(2k+q)^\mu}{2kq+q^2}[F(k+q)-F(k)]                               \nonumber\\
&\times&\frac{1}{D_\pi(k)}\frac{1}{\slashed{p}-\slashed{k}-m_\Delta}S_{\sigma\rho}F(k)(k_\rho+z\gamma_\rho\slashed{k})u(p).
\end{eqnarray}


\end{document}